\documentclass[fp,twocolumn]{jpsj3}
\usepackage{txfonts}
\usepackage[dvipdfmx]{graphicx}
\usepackage{bm,braket}
\usepackage{color}
\usepackage{cite}
\title{
Electronic States and Energy Dissipations of Vortex Core in Pure FeSe Single Crystals Investigated by Microwave Surface Impedance Measurements
}

\author{
Tatsunori Okada$^{1,2}$\thanks{tatsunori.okada.d8@tohoku.ac.jp, cmaeda@g.ecc.u-tokyo.ac.jp}, 
Yoshinori Imai$^{1,3}$,
Takahiro Urata$^{3}$\thanks{Present address: Department of Materials Physics, Nagoya University, Nagoya, Aichi 464-8603, Japan},
Yoichi Tanabe$^{3}$\thanks{Present address: Department of Applied Science, Okayama University of Science, Okayama, Okayama 700-0005, Japan},
Katsumi Tanigaki$^{3,4}$,
and Atsutaka Maeda$^{1*}$
}

\inst{
$^{1}$Department of Basic Science, University of Tokyo, Meguro-ku, Tokyo 153-8902, Japan
\\
$^{2}$Institute for Materials Research, Tohoku University, Sendai, Miyagi 980-8577, Japan
\\
$^{3}$Department of Physics, Graduate School of Science, Tohoku University, Sendai, Miyagi 980-8578, Japan
\\
$^{4}$WPI Advanced Institute for Materials Research, Tohoku University, Sendai, Miyagi 980-8577, Japan
} 

\abst{
In order to clarify electronic states and energy dissipations due to a motion of a vortex core in pure FeSe, which is a candidate superconductor possessing a super-clean core, we measured the microwave surface impedance of pure FeSe single crystals under finite magnetic fields.
From the magnetic-field dependence of the flux-flow resistivity, we found that a barometer of electronic states inside the vortex core $\omega_{0}\tau_{\rm core}$ is $1\pm0.5$, suggesting that the vortex core of pure FeSe is in the moderately clean regime contrary to the expectation of the super-clean core.
We also found that the mean-free path inside the vortex core is suppressed at the distance of the order of the core radius.
Based on observed results and previous reports, we discussed possible origins of rather small $\omega_{0}\tau_{\rm core}$ value in terms of the multiple-bands nature of FeSe and additional mechanisms producing extra energy dissipations specific to the vortex core in motion.
}
\begin{document}
\maketitle
\section{Introduction}
\begin{figure}[h]
	\centering
	\includegraphics[width=\hsize]{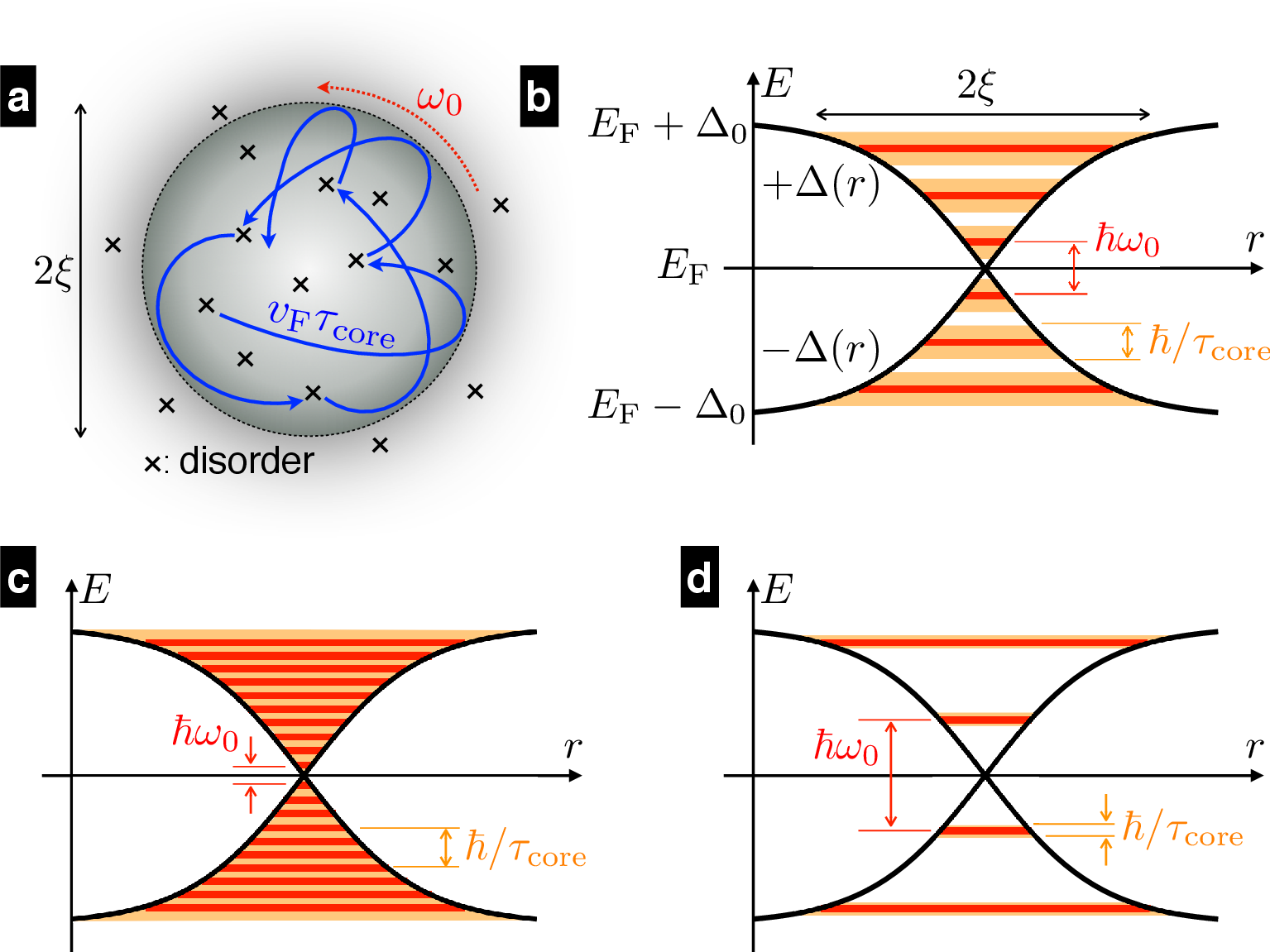}
	\caption{(Color online) 
		Schematics of vortex-core bound states.
		(a) Quasiparticles travel in the vortex core with the mean-free path of $v_{\rm F}\tau_{\rm core}$ and the precession angle of $\omega_{0}$.
		Energy spectrums (b) in the case of moderately clean core ($\omega_{0}\tau_{\rm core}\simeq1$), (c) in conventional SCs ($\omega_{0}\tau_{\rm core}\ll1$; dirty core), and (d) expected in pure FeSe ($\omega_{0}\tau_{\rm core}\gg1$; super-clean core).
	}
	\label{fig1}
\end{figure}

The discovery of superconductivity in LaFeAsO$_{1-x}$F$_x$ with a critical temperature of $T_{\rm c}=26\ {\rm K}$ \cite{Kamihara2008} triggered energetic researches on iron-based superconductors (FeSCs).
In order to explore the origin of superconductivity and/or to clarify their potential for superconducting applications, experimental investigations on many physical properties of FeSCs have been carried out so far.
The flux-flow resistivity, which reflects energy dissipations induced by quasiparticles bound inside and vicinity of the vortex core (Fig.\ref{fig1}a), is one of such physical properties.
Since the vortex core is formed by superconducting-gap function $\Delta$, the flux-flow resistivity $\rho_{\rm f}$ contains information on electronic states inside the vortex core and on the superconducting-gap structure related to the pairing mechanism.

As schematically shown in Fig.\ref{fig1}b, vortex-core-bound states which are known as Caroli-deGennes-Matricon (CdGM) modes \cite{Caroli1964} are characterized by two energy scales; an energy spacing $\hbar\omega_{0}=\Delta_{0}^{2}/E_{\rm F}$ and a width of each levels $\hbar/\tau_{\rm core}$, where $\Delta_{0}$ is a superconducting gap far from the vortex core, $E_{\rm F}$ a Fermi energy, $\tau_{\rm core}$ a scattering time of quasiparticles inside the vortex core.
The ratio of the spacing to the width $\omega_{0}\tau_{\rm core}$ is used as a barometer of electronic states inside the vortex core, which can be classified into (i) dirty core ($\omega_{0}\tau_{\rm core}\ll1$), (ii) moderately clean core ($\omega_{0}\tau_{\rm core}\sim1$), or (iii) super-clean core ($\omega_{0}\tau_{\rm core}\gg1$) \cite{Blatter1994,Golosovsky1996}.
Since conventional SCs possess a small $\Delta_{0}$ ($\lesssim10\ {\rm K}$) and a large $E_{\rm F}$ ($\sim10^{4}\ {\rm K}$), energy levels with a fine spacing of $\hbar\omega_{0}\sim10^{-4}\Delta_{0}$ are formed below $\Delta_{0}$ (Fig.\ref{fig1}c).
Thus, CdGM modes in conventional SCs look like almost continuous energy spectrum \cite{Hess1989}, and the electronic state inside the vortex core results in the dirty core ($\omega_{0}\tau_{\rm core}\sim0.01$).
In contrast, high-$T_{\rm c}$ SCs, such as copper oxides and iron pnictides/chalcogenides, possess larger $\Delta_{0}$ corresponding to higher $T_{\rm c}$, and $\omega_{0}\tau_{\rm core}$ in those materials has been expected to be much larger than that in conventional SCs.
An early study on YBa$_2$Cu$_3$O$_{7-\delta}$ single crystals by using a microwave technique \cite{Matsuda1994} claimed $\omega_{0}\tau_{\rm core}>1$, but the evaluation of $\omega_{0}\tau_{\rm core}$ is not accurate because $\omega_{0}\tau_{\rm core}$ was estimated solely from the resistive response of the sample by using a relatively low frequency.
Recent studies on single crystals of YBa$_2$Cu$_3$O$_{7-\delta}$ \cite{Tsuchiya2001,Golosovsky1996,Maeda2007_YBCO}, Bi$_2$Sr$_2$CaCu$_2$O$_{8+\delta}$ \cite{Maeda2001}, La$_{2-x}$Sr$_x$CuO$_4$ \cite{Maeda2007_LSCO}, LiFeAs$_{1-x}$P$_x$ (with $x = 0$ \cite{Okada2012,Okada2020_FluxFlow} and 0.03 \cite{Okada2013_P-Li111}), NaFe$_{0.97}$Co$_{0.03}$As \cite{Okada2013_Co-Na111}, FeSe$_{0.4}$Te$_{0.6}$ (synthesized by a melt-growth method \cite{Okada2015}) by measuring both of the resistive and reactive microwave responses revealed that electronic states in the vortex core of these materials are still in the moderately clean regime ($\omega_{0}\tau_{\rm core}=0.1-0.5$).
Therefore, SCs with the super-clean core have not been confirmed, and the dynamics and the dissipation mechanism of super-clean core have not been clarified yet.

Although the flux-flow phenomena is one of the fundamental topic of superconductivity discovered more than fifty-years ago, the nature of magnetic vortex in motion has not been fully understood.
Indeed, the origin of the force driving magnetic vortices, namely the Lorentz (electromagnetic) force\cite{Bardeen1965,Kopnin_book2001} and/or the Magnus (hydrodynamic) force\cite{Nozieres1966,Ao1993}, is still under the debate\cite{Kato2016}.
Furthermore, our experimental results reported previously \cite{Tsuchiya2001,Maeda2007_YBCO,Maeda2001,Maeda2007_LSCO,Okada2012,Okada2013_P-Li111,Okada2013_Co-Na111,Okada2015} suggest that a novel mechanism of energy dissipation, which has not been elucidated yet, may exist around the vortex core {\it in motion}.
This indicates that the interpretation about the flux-flow phenomenon so far may be missing important factors for understanding the flux-flow phenomena.
Therefore, experimental investigations on the flux-flow phenomena is an important issue for understanding the superconductivity, and the elucidation of the motion and dissipation mechanism of the super-clean core is expected to bring valuable knowledge.

In this article, we focus on FeSe single crystals synthesized by a chemical vapor deposition with a KCl-AlCl$_3$ flux \cite{Bohmer2013} (``pure FeSe").
Pure FeSe shows a dc resistivity of $\rho_{\rm dc}(T_{\rm c})\sim10\ {\mu\Omega}{\rm cm}$ \cite{Bohmer2013,Kasahara2014}, which is one-order of magnitude smaller than that in FeSe$_{1-x}$Te$_{x}$ grown by a melt-growth method.
Hence, the quasiparticle scattering time inside the vortex core $\tau_{\rm core}$ is expected to be long.
Furthermore, FeSe$_{1-x}$Te$_x$ system is known to have a very small $E_{\rm F}$ comparable to $\Delta_{0}$ \cite{Lubashevsky2012,Maletz2014,Okazaki2014,Kasahara2014,Hanaguri2019}.
From a viewpoint of the vortex core, $\Delta_{0}/E_{\rm F}\simeq1$ can be interpreted as a quantum-limit core \cite{Hayashi1998_QuantumLimit}, where the energy spacing $\hbar\omega_{0}=\Delta_{0}^{2}/E_{\rm F}$ is comparable to $\Delta_{0}$ (Fig.\ref{fig1}d).
Indeed, a Friedel-type spatial oscillation of CdGM modes \cite{Hayashi1998_QuantumLimit} was observed by STS measurements on FeSe thin films and single crystals \cite{Song2011,Hanaguri2019}.
By combining the high-purity and the quantum-limit-core, pure FeSe is the most promising candidate for the super-clean core.

To elucidate electronic states and energy dissipations regarding the vortex core and to see if the super-clean core is realized in pure FeSe, we investigated the flux-flow resistivity of pure FeSe single crystals on the basis of microwave surface impedance measurements under finite magnetic fields.
\section{Experimental}
\subsection{Synthesis and transport characteristics}
\begin{figure}[t]
	\centering	
	\includegraphics[width=\hsize]{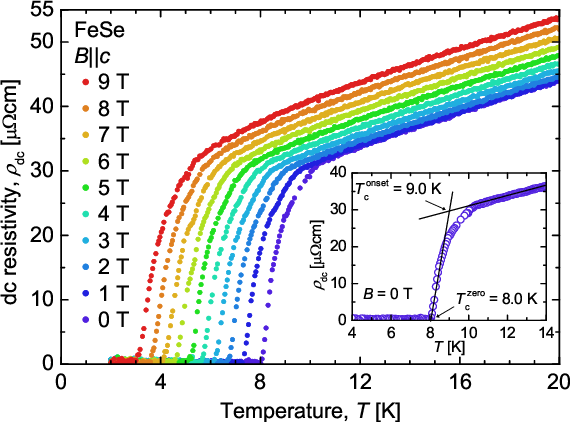}
	\caption{(Color online) 
		Temperature dependence of dc resistivity of pure FeSe single crystals under magnetic fields up to 9 T applied along the $c$ axis.
		The inset is an enlarged plot of $\rho_{\rm dc}(T)$ at $B=0\ {\rm T}$.
	}
	\label{fig2}
\end{figure}

Pure FeSe single crystals were synthesized by the chemical vapor deposition with the KCl-AlCl$_3$ flux \cite{Bohmer2013,Huynh2014,Urata2016}.
Detailed conditions and procedures of synthesis were reported in Refs. \citen{Huynh2014} and \citen{Urata2016}.
We measured the temperature dependence of the dc resistivity $\rho_{\rm dc}(T)$ under magnetic fields $B$ up to 9 T by using a conventional four-terminals method with a physical properties measurement system (PPMS, Quantum Design, Inc.) (Fig. \ref{fig2}).
$T_{\rm c}^{\rm onset}$ and $T_{\rm c}^{\rm zero}$ under $B=0\ {\rm T}$ are 9.0 K and 8.0 K, respectively.
The residual resistivity evaluated by a linear extrapolation of $\rho_{\rm dc}(14\ {\rm K}\leq T\leq20\ {\rm K},0\ {\rm T})$ to 0 K is about $\rho_{\rm dc}^{\rm res}=18\ {\rm \mu\Omega cm}$, and the residual resistivity ratio defined by ${\rm RRR}=\rho_{\rm dc}(300\ {\rm K},0\ {\rm T})/\rho_{\rm dc}^{\rm res}$ was about 24.
These values indicate that the purity of our FeSe single crystals are higher than that of conventional FeSe$_{1-x}$Te$_x$ grown by the melt-growth method \cite{Okada2015}.

$\rho_{\rm dc}$ increases by applying magnetic fields similarly to previous reports \cite{Bohmer2013,Kasahara2014,Huynh2014}, but observed magnetoresistance is smaller than that in Ref. \citen{Kasahara2014}.
This discrepancy might come from differences in $T_{\rm c}$ and/or $\rho_{\rm dc}(0\ {\rm K},0\ {\rm T})$ values.
The normal-state resistivity $\rho_{\rm dc}(T>T_{\rm c},B)$ was used to determine the magnitude of the microwave surface impedance explained below.
\subsection{Microwave surface impedance measurements}
The flux-flow resistivity was investigated by using a cavity perturbation technique in a microwave region \cite{Klein1993}.
We used two cylindrical cavity resonators made of oxygen-free copper operated in the TE$_{011}$ mode.
Those resonators have resonant characteristics of $(\omega_{\rm blank}/2\pi, Q_{\rm blank})=(19.6\ {\rm GHz}, 6.2\times10^{4})$ and $(43.9\ {\rm GHz}, 2.6\times10^{4})$ under (4.2 K, 0 T) condition, where $\omega_{\rm blank}/2\pi$ and $Q_{\rm blank}$ are the resonant frequency and the quality factor of the resonator without inserting a sample into the resonator.
Pure FeSe single crystals (batch \#1 and \#2) were cut into a rectangular shape with dimensions of $0.4\times0.4\times0.1\ {\rm mm}^3$ and placed at the center of the resonator by mounting on a sapphire rod with a diameter of 1 mm.
External magnetic fields up to 8 T and weak microwave fields were applied parallel to the $c$ axis of the sample, producing a in-plane motion of vortices.
The microwave surface impedance, $Z_{\rm s}=R_{\rm s}-{\rm i}X_{\rm s}$, can be obtained from the shifts of resonant frequency and quality factor between with- and without sample conditions;
\begin{eqnarray}
	R_{\rm s}(T,B)
	\hspace{-0.8em}&=&\hspace{-0.8em}G\left(
		\dfrac{1}{2Q_{\rm sample}(T,B)}
		-\dfrac{1}{2Q_{\rm blank}(T,B)}
	\right),
	\\
	X_{\rm s}(T,B)
	\hspace{-0.8em}&=&\hspace{-0.8em}C-G\left(\dfrac{\omega_{\rm sample}(T,B)-\omega_{\rm blank}(T,B)}{\omega_{\rm blank}(T,B)}\right),
\end{eqnarray}
where $G$ and $C$ are constants depending on geometries of the sample and the resonator.
These factors can be determined by imposing additional conditions of $R_{\rm s}(T>T_{\rm c},B)=X_{\rm s}(T>T_{\rm c},B)=\sqrt{\mu_{0}\omega\rho_{\rm dc}(T>T_{\rm c},B)/2}$ with $\rho_{\rm dc}(T,B)$ shown in Fig. \ref{fig1}.
The surface impedance below $T_{\rm c}$ can be expressed as $Z_{\rm s}(T,B)=-{\rm i}\mu_{0}\omega\tilde{\lambda}(T,B)$ with the complex penetration depth \cite{Coffey1991}
\begin{eqnarray}
	\tilde{\lambda}
	=\lambda\sqrt{\dfrac{1}{1-{\rm i}u}\left(1+{\rm i}\dfrac{\rho_{\rm f}}{\mu_{0}\omega\lambda^{2}}\dfrac{\varepsilon-{\rm i}\omega/\omega_{\rm cr}}{1-{\rm i}\omega/\omega_{\rm cr}}\right)},
\end{eqnarray}
where $\lambda(T)$ is the penetration depth in the zero-field limit, $\rho_{\rm f}(T,B)$ the flux-flow resistivity, $\omega_{\rm cr}(T,B)/2\pi$ the crossover frequency characterizing the crossover from a reactive response ($\omega\ll\omega_{\rm cr}$) to a resistive response ($\omega\gg\omega_{\rm cr}$) \cite{Gittleman1966}.
Dimensionless parameters $\varepsilon$ and $u$ reflect contributions of the flux creep and the normal fluid on $\tilde{\lambda}$.
We set $\varepsilon=0$ and $u=0.4\ (44\ {\rm GHz}), 0.2\ (19\ {\rm GHz})$ for analyzing data measured at 2 K (see Appendix).
Consequently, we can evaluate $\rho_{\rm f}(T,B)$ and $\omega_{\rm cr}(T,B)/2\pi$ of the sample by measuring the real- and imaginary parts of $Z_{\rm s}(T,B)$.
\section{Results and discussions}
\subsection{Microwave surface impedance}
\begin{figure}[h]
	\centering
	\includegraphics[width=\hsize]{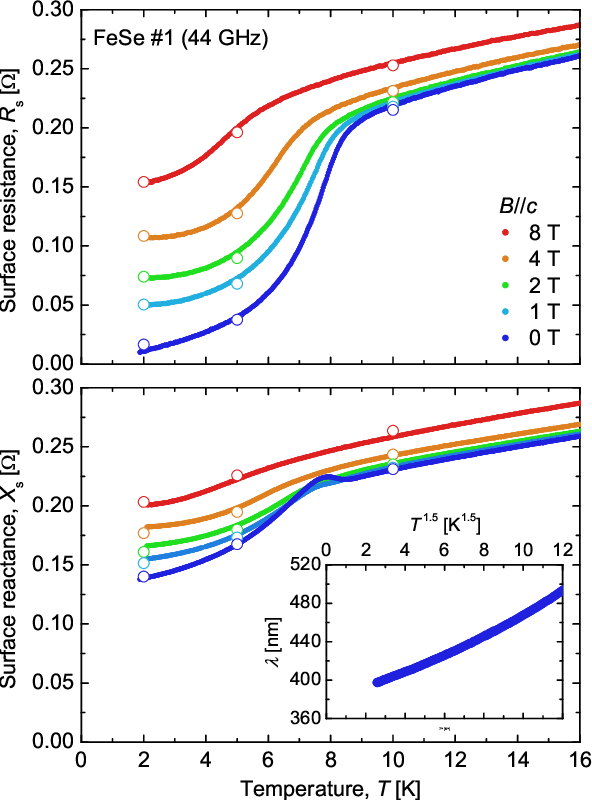}
	\caption{(Color online) 
		Temperature dependence of microwave surface impedance, $Z_{\rm s}=R_{\rm s}-{\rm i}X_{\rm s}$, of pure FeSe single crystal \#1 measured at 44 GHz under fixed external magnetic fields.
		Open circles are those measured by sweeping magnetic fields at constant temperatures.
		The inset shows the penetration depth $\lambda(T)=X_{\rm s}(T,0\ {\rm T})/\mu_{0}\omega$ as a function of $T^{1.5}$.
	}
	\label{fig3}
\end{figure}

Figure \ref{fig3} shows the temperature dependence of microwave surface impedance of pure FeSe \#1.
A good agreement between the data measured by sweeping $T$ and those by sweeping $B$ indicates a uniform distribution of vortices in the sample at least in an effective region where magnetic fields penetrate.

As shown in the inset of Fig. \ref{fig3}, the penetration depth evaluated by $\lambda(T)=X_{\rm s}(T,0\ {\rm T})/\mu_{0}\omega$ increases as $T^{n}$ with an exponent of $n\simeq1.5$ at low-$T$, and its magnitude in the 0 K limit is $\lambda(0\ {\rm K})=377\pm10\ {\rm nm}$.
These values are consistent with $n=1.4$ and $\lambda(0\ {\rm K})\simeq400\ {\rm nm}$ reported in Ref. \citen{Kasahara2014}.
Since $\lambda(T)$ is sensitive to low-energy quasiparticle excitations, the presence of a line-nodal gap \cite{Kasahara2014} or a highly anisotropic nodeless gap \cite{Li2016} has been proposed from the penetration depth point of view.
Unfortunately, the lowest temperature we investigated is rather high for discussing the topology of superconducting gaps, and we do not address it in this paper.

\subsection{Crossover frequency}
\begin{figure}[h]
	\centering
	\includegraphics[width=\hsize]{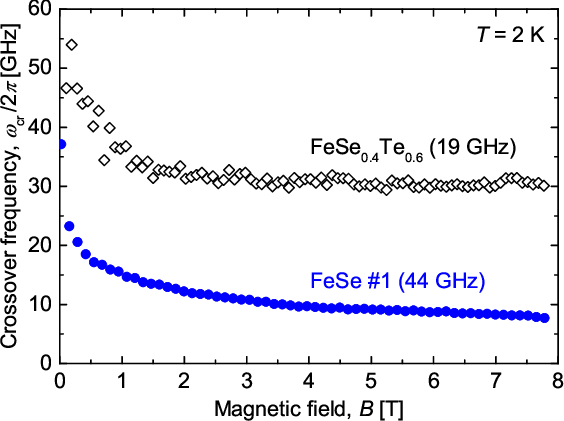}
	\caption{
		(Color online) Magnetic field dependence of crossover frequency, $\omega_{\rm cr}/2\pi$, of pure FeSe single crystal \#1 measured at 2 K (blue circles).
		$\omega_{\rm cr}(2\ {\rm K})/2\pi$ of FeSe$_{0.4}$Te$_{0.6}$ single crystal synthesized by the melt-growth method \cite{Okada2015} is also depicted for comparison as black diamonds.
	}
	\label{fig4}
\end{figure}

Next, we work on the crossover frequency, $\omega_{\rm cr}/2\pi$, which characterizes the crossover between reactive- and resistive responses of the vortex motion \cite{Gittleman1966}.
As shown in Fig. \ref{fig4}, $\omega_{\rm cr}(2\ {\rm K})/2\pi$ of pure FeSe decreases gradually with increasing $B$.
A similar behavior is widely observed in FeSCs and can be understood by weakening of pinning force due to vortex-vortex repulsions.

Regarding the magnitude of the crossover frequency, $\omega_{\rm cr}/2\pi$ of pure FeSe is about 15 GHz under (2 K, 1T) condition, which is lower than that in FeSe$_{0.4}$Te$_{0.6}$ ($\simeq37\ {\rm GHz}$ shown in Fig. \ref{fig4} for comparison) \cite{Okada2015} and in FeSe$_{0.5}$Te$_{0.5}$ thin films ($\simeq21.5\ {\rm GHz}$ at 12 K) \cite{Pompeo2020}.
It is expected that single crystals of pure FeSe have less disorders compared to those of FeSe$_{1-x}$Te$_{x}$ ($x>0$) containing excess Fe atoms and Se-Te substitutions.
It has been clarified that an introduction of disorders to pure FeSe single crystals by proton-irradiation enhances the pinning force \cite{Sun2015}.
Thus, low $\omega_{\rm cr}$ of pure FeSe indicates the high purity of this material.
\subsection{Flux-flow resistivity and electronic state in vortex core}
\begin{figure}[h]
	\centering
	\includegraphics[width=\hsize]{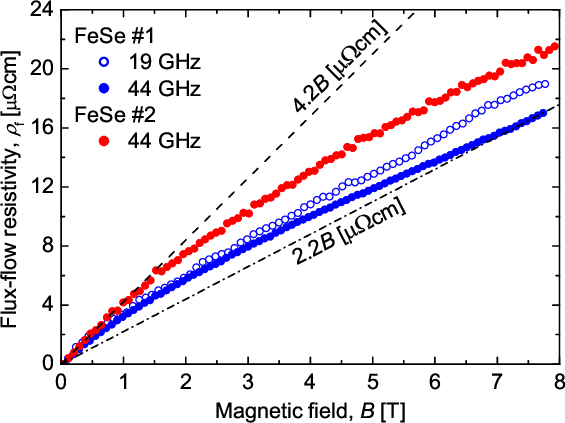}
	\caption{
		(Color online) Magnetic field dependence of flux-flow resistivity, $\rho_{\rm f}(B)$, of pure FeSe measured at $2\ {\rm K}$.
		Dashed and dot-dashed lines are $4.2B$ [$\mu\Omega$cm] and $2.2B$ [$\mu\Omega$cm] as guides for eyes.
	}
	\label{fig5}
\end{figure}

Figure \ref{fig5} shows the magnetic field dependence of the flux-flow resistivity of pure FeSe measured at 2 K.
We found that $\rho_{\rm f}(B)$ increases with $B$ similarly to other FeSCs \cite{Okada2012,Okada2013_P-Li111,Okada2013_Co-Na111,Okada2014_P-Ba122,Okada2015,Okada2020_FluxFlow}.
In particular, a convex upward increase of $\rho_{\rm f}(B)$ resembles to 122-type FeSCs possessing a line-nodal gap \cite{Takahashi2012,Okada2014_P-Ba122,Okada2020_FluxFlow}, being consistent with the presence of nodal lines or small minima in superconducting gaps of FeSe \cite{Kasahara2014,Liu2018}.

According to microwave investigations on $\rho_{\rm f}(B)$ in FeSCs \cite{Okada2012,Takahashi2012,Okada2013_P-Li111,Okada2013_Co-Na111,Okada2014_P-Ba122,Okada2015,Okada2020_FluxFlow}, the initial slope of the flux-flow resistivity normalized by the normal-state resistivity with respect to the normalized magnetic field, $\alpha=\left.(\rho_{\rm f}/\rho_{\rm n})/(B/B_{\rm c2})\right|_{B\rightarrow0}$, reflects the superconducting gap anisotropy and the multiple-bands nature.
As discussed in $\rho_{\rm f}(B)$ of FeSe$_{0.4}$Te$_{0.6}$ single crystal \cite{Okada2015}, the upper critical field in the orbital limit $B_{\rm c2}^{\rm orb}$ should be used for $B_{\rm c2}$.
However, unfortunately, it is difficult to evaluate $B_{\rm c2}^{\rm orb}$ accurately because $B_{\rm c2}(T)$ of FeSe should be strongly influenced by the spin-orbit interaction expected from a large Maki parameter reflecting the large $\Delta/E_{\rm F}$ ratio.
In addition, the remarkable magnetoresistance (Fig. \ref{fig1}) also makes it difficult to evaluate the normal-state resistivity at $B_{\rm c2}^{\rm orb}$ accurately.
Therefore, it is needed to measure $\rho_{\rm f}(B)$ well above $B_{\rm c2}^{\rm orb}$ for evaluating the initial slope $\alpha$, and such measurements under ultra high fields are beyond the scope of this paper.

Now let us evaluate the barometer of electronic states in the vortex core, $\omega_{0}\tau_{\rm core}=\Phi_{0}B/n\pi\hbar\rho_{\rm f}$, from the flux-flow resistivity.
As shown in Fig. \ref{fig5}, obtained $\rho_{\rm f}(B)$ at 2 K below 8 T is in the range of $2.2\ {\rm \mu\Omega cm/T}\leq\rho_{\rm f}(B)/B\leq4.2\ {\rm \mu\Omega cm/T}$.
Measurements on the Shubnikov-de Haas oscillation \cite{Terashima2014} and the Hall resistivity \cite{Huynh2014} in pure FeSe single crystals reported the carrier density of $n=3.1\times10^{20}\ {\rm cm}^{-3}$ and $1.9\times10^{20}\ {\rm cm}^{-3}$, respectively.
By using these numbers, the barometer of electronic states in the vortex core results in $\omega_{0}\tau_{\rm core}=1\pm0.5$.
Some of the authors obtained a similar $\omega_{0}\tau_{\rm core}=|\rho_{\rm fH}/\rho_{\rm f}|$ value \cite{Ogawa2021_unpublished} by measuring the longitudinal- ($\rho_{\rm f}$) and Hall ($\rho_{\rm fH}$) components of the flux-flow resistivity directly with a cross-shaped bimodal cavity operated in a microwave region \cite{Ogawa2021}.
$\omega_{0}\tau_{\rm core}=1\pm0.5$ is the largest among other FeSCs measured in so far; LiFeAs ($0.4\pm0.1$) \cite{Okada2012}, NaFe$_{0.97}$Co$_{0.03}$As ($0.14\pm0.06$) \cite{Okada2013_Co-Na111}, BaFe$_2$(As$_{0.55}$P$_{0.45}$)$_2$ ($0.11\pm0.06$ calculated by using data in Ref. \citen{Okada2014_P-Ba122}), and FeSe$_{0.4}$Te$_{0.6}$ ($0.14\pm0.03$ calculated by using data in Ref. \citen{Okada2015}).
The large $\omega_{0}\tau_{\rm core}$ value is consistent with the high purity and the large $\Delta_{0}/E_{\rm F}$ ratio of pure FeSe, but obtained $\omega_{0}\tau_{\rm core}\simeq1$ is still in the moderately clean regime.

Since $\rho_{\rm dc}(0\ {\rm K},0\ {\rm T})$ of our samples are $3-4$ times larger than that in Ref. \citen{Kasahara2014}, pure FeSe with higher purity may have a $3-4$ times longer scattering time.
However, even in this case, $\omega_{0}\tau_{\rm core}$ is expected to be of the order of unity; the vortex core of pure FeSe is still in the moderately clean regime.
This is inconsistent with the expectation that the super-clean core is realized in pure FeSe.
Here we consider the reason why $\omega_{0}\tau_{\rm core}$ of pure FeSe resulted in about unity below.

One possible origin is a carrier compensation reflecting an existence of hole- and electron-type bands in FeSe \cite{Nakayama2014,Kasahara2014,Phan2017}.
According to Ref. \citen{Kopnin_book2001} in which hole- and electron-like contributions to the ohmic and Hall components of the flux-flow resistivity were taken into account, a cancellation due to a carrier compensation similar to the normal-state Hall resistivity is proposed in the flux-flow Hall resistivity, $\rho_{\rm fH}$.
In this case, the absolute value of the tangent of net flux-flow Hall angle,
\begin{eqnarray}\nonumber
	\omega_{0}\tau_{\rm core}
	=\left|\dfrac{\rho_{\rm fH}}{\rho_{\rm f}}\right|
	\hspace{-0.8em}&\simeq&\hspace{-0.8em}\left|\dfrac{\rho_{\rm fH}^{\rm hole}}{\rho_{\rm f}}+\dfrac{\rho_{\rm fH}^{\rm elec}}{\rho_{\rm f}}\right|
	\\
	\hspace{-0.8em}&=&\hspace{-0.8em}\left|(\omega_{0}\tau_{\rm core})^{\rm hole}-(\omega_{0}\tau_{\rm core})^{\rm elec}\right|,
\end{eqnarray}
is expected to be smaller than $(\omega_{0}\tau_{\rm core})^{\rm hole}$ and $(\omega_{0}\tau_{\rm core})^{\rm elec}$ alone, where $(\omega_{0}\tau_{\rm core})^{\rm hole/elec}$ is $\omega_{0}\tau_{\rm core}$ due to hole/electron bands.
Along this speculation, the net $\omega_{0}\tau_{\rm core}$ can be $\omega_{0}\tau_{\rm core}\lesssim1$ even when $(\omega_{0}\tau_{\rm core})^{\rm hole}$ and/or $(\omega_{0}\tau_{\rm core})^{\rm elec}$ are in the super-clean-core regime, $(\omega_{0}\tau_{\rm core})^{\rm hole/elec}\gg1$, as expected in pure FeSe.
A remarkable reduction of the net carrier density observed by the normal-state Hall resistivity measurements in pure FeSe \cite{Huynh2014} may support this scenario.

An uncertain point of the carrier compensation scenario is whether contributions of carriers to $\rho_{\rm fH}$ can be considered to be the same as those to the Hall resistivity in the normal state, $\rho_{\rm H}$.
For instance, $\rho_{\rm fH}$ of several single-band SCs (V \cite{Noto1976}, Bi$_2$Sr$_2$CaCu$_2$O$_{8+\delta}$, and ErBa$_2$Cu$_3$O$_{7-\delta}$ \cite{Iye1989}) possesses the sign opposite to $\rho_{\rm H}$.
This indicates that additional contributions affect the flux-flow phenomena, and a charging of the vortex core has been proposed as a candidate for such additional effects \cite{Khomskii1995,Kato1999,Kohno2016}.
Therefore, it is difficult to conclude that observed $\omega_{0}\tau_{\rm core}\simeq1$ is induced solely by the carrier compensation scenario at present, and other origins, including the charged vortex core, may limit $\omega_{0}\tau_{\rm core}$ value.
\subsection{Quasiparticle scattering time}
\begin{figure}[h]
	\centering
	\includegraphics[width=\hsize]{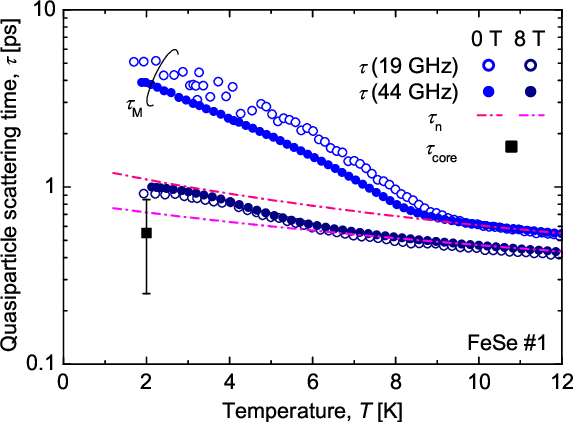}
	\caption{
		(Color online) Temperature dependence of quasiparticle scattering time, $\tau$, of pure FeSe single crystal \#1 in different conditions.
		 Blue circles are $\tau$ evaluated from the surface impedance, and those measured at 0 T are $\tau$ in the Meissner state, $\tau_{\rm M}$.
		Dot-dashed curves are $\tau$ in the normal state, $\tau_{\rm n}$, which are evaluated by extrapolating $\tau(T>T_{\rm c})$ to $T<T_{\rm c}$.
		The black square is $\tau$ in the vortex core, $\tau_{\rm core}$.
	}
	\label{fig6}
\end{figure}

Now let us focus on $\tau_{\rm core}$ to explore another candidate affecting on $\omega_{0}\tau_{\rm core}$.
STS measurements at the center of the vortex core in pure FeSe single crystals \cite{Hanaguri2019} and thin films \cite{Song2011} observed a differential conductance peak of $V_{\rm peak}\simeq0.6\ {\rm meV}$.
By assuming this peak is the lowest CdGM level and by using $\omega_{0}\tau_{\rm core}=1\pm0.5$ at 2 K, we have $\tau_{\rm core}=0.55\pm0.3\ {\rm ps}$.
In Fig. \ref{fig6}, we compared $\tau_{\rm core}$ with quasiparticle scattering time in the Meissner state corresponding to the outside of the vortex core, $\tau_{\rm M}$, and that in the normal state, $\tau_{\rm n}$.
We let $\tau_{\rm M}(T)$ be $\tau(T,0\ {\rm T})$ obtained from $Z_{\rm s}(T,0\ {\rm T})$ data with Eq. \eqref{eq:tau}.
$\tau_{\rm n}(T,B)$ was evaluated by fitting $\tau(T>T_{\rm c},B)$ with the inverse of a $T$-linear function, $[a(B)+b(B)T]^{-1}$, and by extrapolating it into $T<T_{\rm c}$.
As shown in Fig. \ref{fig6}, obtained $\tau_{\rm core}$ is an order of magnitude shorter than $\tau_{\rm M}$ at low $T$.
A similar suppression of $\tau_{\rm core}$ compared to $\tau_{\rm M}$ has been reported in many SCs, such as YBa$_2$Cu$_3$O$_{7-\delta}$ \cite{Tsuchiya2001}, Bi$_2$Sr$_2$CaCu$_2$O$_{8+\delta}$ \cite{Maeda2001}, La$_{2-x}$Sr$_x$CuO$_4$ \cite{Maeda2007_LSCO}, Y$_2$C$_3$ \cite{Akutagawa2008}, and LiFeAs$_{1-x}$P$_x$ \cite{Okada2012,Okada2013_P-Li111}.
Furthermore, $\tau_{\rm core}$ is slightly shorter than $\tau_{\rm n}(8\ {\rm T})$.
As mentioned before, we do not know the exact upper critical field and normal-state resistivity at 2 K, but $\tau_{\rm core}$ is likely to be comparable to $\tau_{\rm n}(B_{\rm c2})$ even though the energy spectrum inside the vortex core differs from the normal state continuum.
This suggests that energy dissipations are specifically enhanced in the vortex core.

By using the Fermi velocity of $v_{\rm F}\simeq3.5\times10^{4}\ {\rm m/s}$, which we evaluated from a band structure \cite{Kasahara2014} with a parabolic approximation, the mean-free path inside- and outside of the vortex core results in $v_{\rm F}\tau_{\rm core}(2 K)=19\pm10\ {\rm nm}$ and $v_{\rm F}\tau_{\rm M}(2 K)=150\pm50\ {\rm nm}$, respectively.
If we use the characteristic field of $B_{0}(2\ {\rm K})\simeq12.7\ {\rm T}$ \cite{Terashima2014} as the lower boundary of the upper critical field, the coherence length becomes $\xi(2\ {\rm K})\leq5\ {\rm nm}$.
Thus, pure FeSe holds the magnitude relationship of $v_{\rm F}\tau_{\rm M}(2\ {\rm K})\gg v_{\rm F}\tau_{\rm core}(2\ {\rm K})\gtrsim\xi(2\ {\rm K})$.
A similar relationship can be seen in cuprates with a highly anisotropic line-nodal gap; YBa$_2$Cu$_3$O$_{7-\delta}$ \cite{Tsuchiya2001}, Bi$_2$Sr$_2$CaCu$_2$O$_{8+\delta}$ \cite{Maeda2001}, La$_{2-x}$Sr$_x$CuO$_4$ \cite{Maeda2007_LSCO}.
In contrast, multiple-bands SCs with nodeless gaps with a moderate anisotropy (LiFeAs \cite{Okada2012} and Y$_2$C$_3$ \cite{Akutagawa2008}) show a different relationship; $v_{\rm F}\tau_{\rm M}\gg\xi\gtrsim v_{\rm F}\tau_{\rm core}$.
The discrepancy in the relationship of $v_{\rm F}\tau_{\rm core}$ and $\xi$ between line-nodal SCs and nodeless SCs probably relates to quasiparticles in node directions leaked out of the vortex core.
In the case of pure FeSe, line nodes \cite{Song2011,Kasahara2014} or finite but small minima ($\lesssim2\ {\rm K}$) of \cite{Bourgeois-Hope2016,Teknowijoyo2016,Li2016} in a superconducting gap have been reported, and quasiparticles along line nodes or deep minima are expected to contribute $\rho_{\rm f}(B)$ at 2 K.
Thus, the fact that pure FeSe holds $v_{\rm F}\tau_{\rm core}\gtrsim\xi$ similar to cuprates is consistent with the gap structure of FeSe. 

In both cases of nodal- and nodelss SCs, it is clear that motions of quasiparticles are suppressed at the distance of the order of core radius; $v_{\rm F}\tau_{\rm core}\sim\xi$.
This indicates the importance of physics specific to the vortex-core boundary to the enhancement of energy dissipations inside the vortex core.
There are some theoretical predictions, which are expected to produce extra energy dissipations inside the vortex core, such as an interplay between a collective motion of the order parameter and CdGM modes \cite{Eschrig1999} and a Landau-Zener tunneling process between CdGM modes \cite{Hayashi1998_SpectralFlow}.
It is unclear that observed energy dissipations can be explained by those theoretical models in a quantitive manner.
There is the possibility that a noble physical mechanism plays a role for extra energy dissipations inside the vortex core in motion.
Therefore, further investigations from theoretical- and experimental aspects are needed to elucidate energy dissipations specific to the vortex core.
From an experimental viewpoint, research on carrier density and/or frequency dependence of the flux-flow resistivity may provide useful information on the dissipation mechanism in the vortex core.
From a theoretical point of view, a microscopic model for the flux-flow phenomena taking account of the multiple-bands nature and the charged vortex core may give a clue to understand our results.
\section{Conclusion}
In order to clarify electronic states and energy dissipations due to the motion of the vortex core in pure FeSe single crystal, which is expected to possess the super-clean core, we investigated the microwave surface impedance in the zero-field limit and under finite magnetic fields.
From the magnetic field dependence of the flux-flow resistivity, we found that the barometer of the electronic state inside the vortex core is $\omega_{0}\tau_{\rm core}=1\pm0.5$.
This suggests that the vortex core of pure FeSe is in the moderately clean regime, which is inconsistent with the expectation that the super-clean core is realized in pure FeSe.
We also found that the mean-free path inside the vortex core is suppressed at the distance of the order of the core radius similarly to cuprates, LiFeAs, and Y$_{2}$C$_{3}$.
We discussed possible origins of rather small $\omega_{0}\tau_{\rm core}$ in terms of the multiple-bands nature and mechanisms producing extra energy dissipations specific to the vortex core.
However, it is not clear that observed $\omega_{0}\tau_{\rm core}$ can be explained by these mechanisms in a quantitative manner, and there is the possibility that novel mechanisms specific to the vicinity of the vortex core play a crucial role for extra energy dissipations.
\section*{Acknowledgment}
\begin{acknowledgment}
T.O. would like to thank Dr. Tetsuo Hanaguri and Dr. Yusuke Masaki for fruitful discussions on vortex-core states in pure FeSe.
This research was partially supported by JSPS KAKENHI (Grant-in-Aid for JSPS Fellows:15J09645 and Early-Career Scientists:18K13783, 21K14192 for T.O.) and (Grant-in-Aid for JSPS Fellows:14J06798 for T.U.).
\end{acknowledgment}
\section*{Author contributions}
T.O., Y.I., and A.M. designed the study.
T.U., Y.T., and K.T. synthesized pure FeSe single crystals and carried out preliminary measurements on transport properties.
T.O. performed transport- and microwave measurements, analyzed and interpreted the data in discussion with Y.I. and A.M.
T.O., Y.I., and A.M. wrote the manuscript with reflecting critical comments from all authors.
\appendix
\section{Effect of flux creep and normal fluid components on complex penetration depth}
\begin{figure}[h]
	\centering
	\includegraphics[width=\hsize]{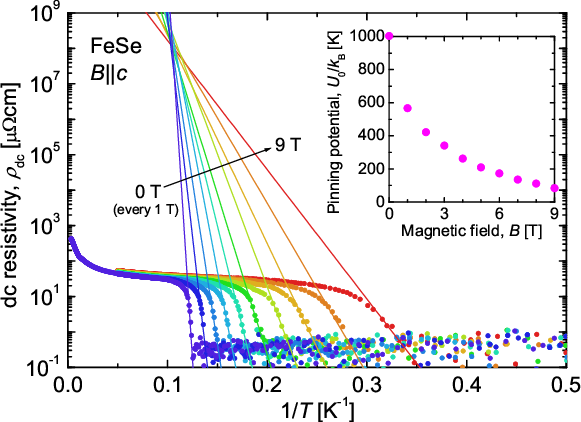}
	\caption{
		(Color online) Arrhenius plot of dc resistivity of FeSe single crystal with magnetic fields up to 9 T.
		Solid lines are fitted results by a simple Arrhenius relation, $\rho_{\rm dc}(T,B)=\rho_{\rm dc}^{0}(B)\exp{\left[-U_{0}(B)/k_{\rm B}T\right]}$.
		The inset is evaluated pinning potential $U_{0}$ as a function of magnetic fields.
	}
	\label{figA1}
\end{figure}

According to Ref. \citen{Coffey1991}, effects of the flux creep on the complex penetration depth is parameterized by $\varepsilon=1/I_{0}^{2}(U_{0}/2k_{\rm B}T)$, where $I_{0}$ is the zero-th order modified Bessel function of the first kind.
By fitting the dc resistivity of pure FeSe single crystal with a simple Arrhenius relation, $\rho_{\rm dc}(T,B)=\rho_{\rm dc}^{0}(B)\exp{\left[-U_{0}(B)/k_{\rm B}T\right]}$ (Fig. \ref{figA1}), we obtained the magnetic-field dependence of the pinning potential as shown in the inset.
$U_{0}(B)/k_{\rm B}$ exceeds 100 K even under 9 T.
This leads to $\varepsilon(2\ {\rm K},9\ {\rm T})\simeq3\times10^{-20}$, which is much smaller than $\omega/\omega_{\rm cr}$ ratio (Fig. \ref{fig4}).
Thus, effects of the flux creep on the complex penetration depth is negligible.

\begin{figure}[h]
	\centering
	\includegraphics[width=\hsize]{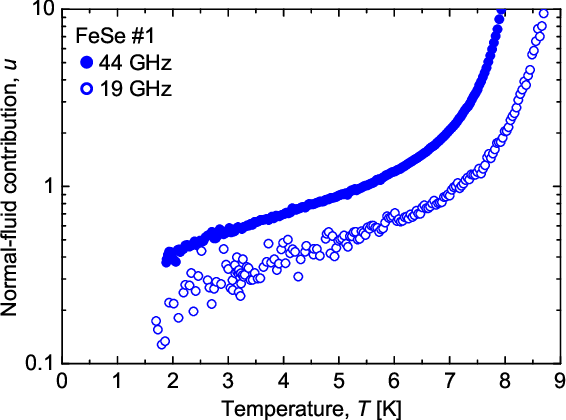}
	\caption{
		(Color online) Temperature dependence of normal-fluid contribution to complex penetration depth, $u=\omega\tau(1-f_{\rm s})/f_{\rm s}$.
	}
	\label{figA2}
\end{figure}

Next we work on normal-fluid contributions to the complex penetration depth, $u=\omega\tau(1-f_{\rm s})/f_{\rm s}$ \cite{Coffey1991}, where $f_{\rm s}$ is the super-fluid density fraction and $\tau$ is the quasiparticle scattering time.
By combining the microwave complex conductivity, $\sigma_{1}+{\rm i}\sigma_{2}=-{\rm i}\mu_{0}\omega/Z_{\rm s}^{2}$, with a two-fluid model, $f_{\rm s}$ and $\tau$ can be calculated by
\begin{eqnarray}
	f_{\rm s}
	\hspace{-0.8em}&=&\hspace{-0.8em}\mu_{0}\omega\lambda^{2}(0\ {\rm K})\sigma_{2}-\dfrac{\left[\mu_{0}\omega\lambda^{2}(0\ {\rm K})\sigma_{1}\right]^{2}}{1-\mu_{0}\omega\lambda^{2}(0\ {\rm K})\sigma_{2}},
	\label{eq:fs}
	\\
	\omega\tau
	\hspace{-0.8em}&=&\hspace{-0.8em}\dfrac{\mu_{0}\omega\lambda^{2}(0\ {\rm K})\sigma_{1}}{1-\mu_{0}\omega\lambda^{2}(0\ {\rm K})\sigma_{2}}.
	\label{eq:tau}
\end{eqnarray}
Evaluated $u$ and $\tau$ are plotted in Fig. \ref{figA2} and Fig. \ref{fig6}.
We set $u=0.4$ and 0.2 for analyzing data measured at 2 K with 44 GHz and 19 GHz, respectively.

\bibliography{210425Reference}


\end{document}